# 'MOHAWK' : a 4000-fiber positioner for DESpec

Will Saunders[1], Greg Smith, James Gilbert, Rolf Muller, Michael Goodwin,

Nick Staszak, Jurek Brzeski, Stan Miziarski, Matthew Colless

Australian Astronomical Observatory, Sydney NSW 1710, Australia

**ABSTRACT**

We present a concept for a 4000-fibre positioner for DESpec, based on the Echidna 'tilting spine' technology. The DESpec focal plane is 450mm across and curved, and the required pitch is ~6.75mm. The size, number of fibers and curvature are all comparable with various concept studies for similar instruments already undertaken at the AAO, but present new challenges in combination. A simple, low-cost, and highly modular design is presented, consisting of identical modules populated by identical spines. No show-stopping issues in accommodating either the curvature or the smaller pitch have been identified, and the actuators consist largely of off-the-shelf components. The actuators have been prototyped at AAO, and allow reconfiguration times of ~15s to reach position errors 7 microns or less. Straightforward designs for metrology, acquisition, and guiding are also proposed. The throughput losses of the entire positioner system are estimated to be ~15%, of which 6.3% is attributable to the tilting-spine technology.

## 1. INTRODUCTION

DESpec is a proposal to undertake a ~5-year spectroscopic survey of ~7M galaxies selected from the Dark Energy Survey (DES)[2], and covering the same area of sky [2]. The aim is use Baryon Acoustic Oscillations and Redshift Space Distortions to constrain models for Dark Energy much more tightly than can be done from the imaging survey alone [3]. DESpec would reuse most of the DES wide field corrector optics, together with a new fiber positioner and spectrographs.

DESpec will have a ~450mm diameter curved focal surface, with ~8m radius of curvature, illuminated at ~F/2.9. These parameters are fixed by the telescope, the existing WFC optics, and the requirement for telecentricity at the focal surface. At least 4000 actuators are required, which means the pitch (the distance between actuator centers) must be 6.75mm or less. Because the required pitch is so small, the currently preferred positioner concept is based on the AAO's 'tilting spine' technology, which is inherently compact and simple. This technology has already been implemented for the FMOS-Echidna instrument for Subaru, which has 400 actuators on a 7.2mm pitch (Figure 1) [4].

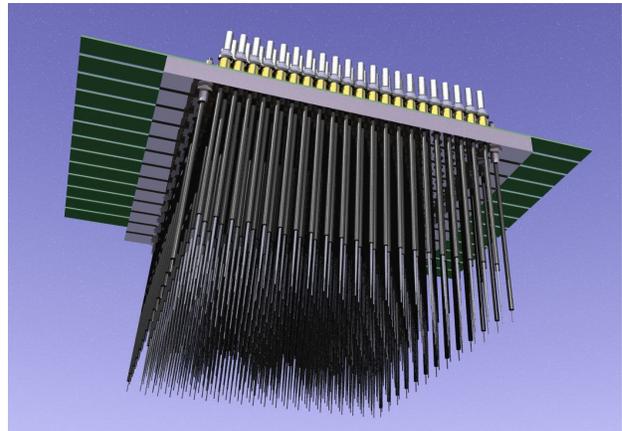

**Figure 1. The FMOS-Echidna positioner for Subaru**

The Echidna actuators consist of two composite parts – a 'base' and a 'spine' (e.g. Figure 5). The base consists of a piezoceramic tube with four electrodes in quadrants on the outside, cemented into the module, together with a cup glued on top which also contains a small ring magnet. The spine consists of a tube containing the fiber, which passes through a steel ball, which is held magnetically in the cup. When a suitable sawtooth waveform is applied to the piezoceramic tube, the resulting stick-slip friction between cup and ball allows the spine to be tilted to any desired position.

---

[1] will@aao.gov.au

[2] DES is a 5-year multiband imaging survey over 5000 deg$^2$, using the Blanco telescope together with a new 2.2° Wide Field Corrector and camera (DECam)[1], and now in its commissioning phase. The aim is to constrain Dark Energy via Weak Lensing and Baryon Acoustic Oscillations.

The actuators are mounted in double rows on 'modules'. Each module consists of a stainless steel bar drilled with holes to mount the actuators, together with a multilayer circuit board bringing power to the actuators (Figure 2). Both actuators and modules are identical and interchangeable.

The tilting spine design has various intrinsic features:

1.  Repositioning require multiple iterations to reach adequate accuracy, but is simultaneous for all fibers[3].

2.  Each actuator has a fixed 'patrol radius' within which it can be positioned. The design allows a reasonable patrol radius, at least as large as the pitch between actuators. This gives at least three-fold covering of the focal surface (Figure 3).

3.  The spines allow a very small minimum separation between targets, ~0.5mm.

4.  The fiber tip is only perfectly in focus at a single radius (Figure 4), causing a small increase in aperture losses at other radii.

5.  The tilting spine design introduces a telecentricity error when positioned away from the home position (Figure 4), resulting in a loss of etendue.

The losses from (4) and (5) depend inversely on spine length, so the longest spines possible are preferred. They can also be traded off against each other (by setting the spines to be in focus near their maximum patrol radius), to give more nearly uniform throughput at all radii.

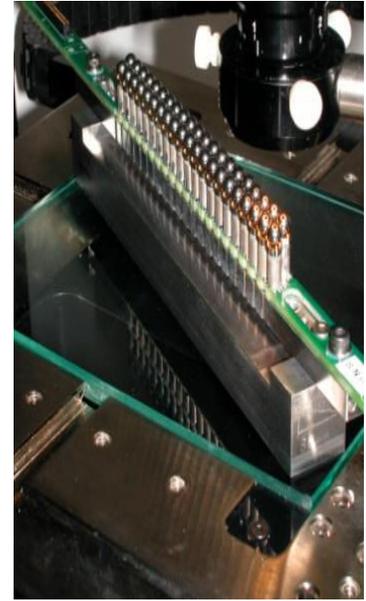

**Figure 2. A single Echidna module**

Extensive further development and prototyping of the tilting-spine design has been undertaken for the WFMOS and WFMOS-A concept designs [10,5], for similar but much larger positioners for Gemini, Subaru, and for the AAT. During that development, the design has been significantly improved. The primary improvements are (a) a very great simplification of the design (Figure 5) (b) the resolution of the issue for Echidna of magnetic cross-talk between close-spaced actuators, and (c) simultaneous metrology of the entire instrument, allowing fully parallel positioning. A complete prototype module exists for the WFMOS design (Figure 6).

The additional changes needed for DESpec are (a) to reduce the pitch to 6.75mm, and (b) to accommodate the spherical focal surface[4]. It was also taken as mandatory to have a simple, modular and low cost and risk design.

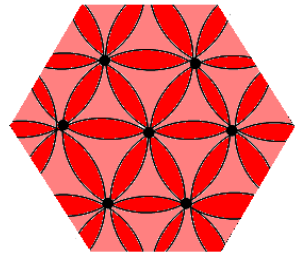

**Figure 3. Schematic view of the patrol areas of adjacent actuators, when the patrol radius is equal to the pitch. The pink areas can be reached by 3 different actuators, while the red areas can be reached by 4 (and the black home positions by 7).**

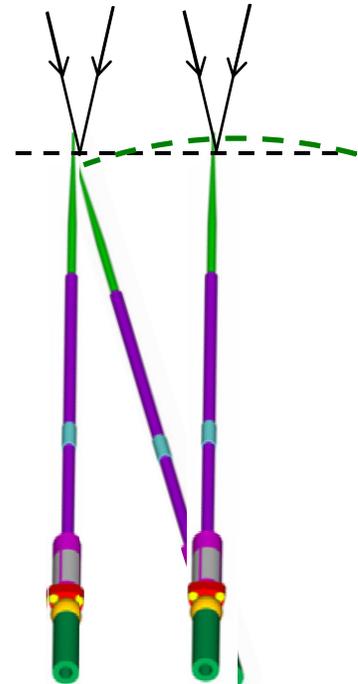

**Figure 4. Schematic drawing of adjacent Echidna spines, showing the focus change with tilt and the telecentricity error at large tilt angle. Black line is focal plane, green line is locus of spine tip.**

---

[3] Positioning is not fully parallel for Echidna, but this is due to limitations in the metrology rather than the positioner.
[4] The WFMOS design for Gemini also followed a curved focal surface, but this was achieved by using spines of adjustable length within a mounted the appropriate tilt angles in a straight module. In the interests of simplicity, scaleability and modularity, we decided for DESpec to use curved modules, with identical spines always mounted at right-angles to the face of module.

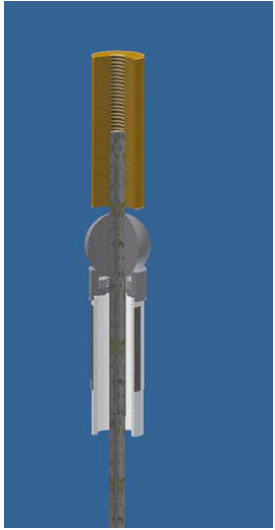
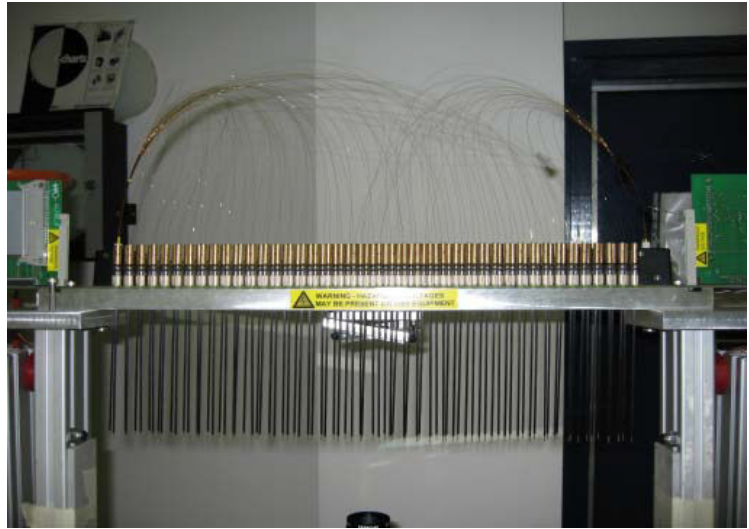

**Figure 5. WFMOS-A actuator and spine**

**Figure 6. Prototype complete WFMOS module**

## 2. MOHAWK STRAWMAN DESIGN

The starting point was the WFMOS-A design. The actuator design for WFMOS-A is already greatly simplified from Echidna. There are just 3 parts in the actuator base and 4 in the spine, and of these 7 parts, 4 are off-the-shelf, 2 require only simple machining, and only one (the cup under the ball bearing) is a significant machining effort (7 surfaces). The entire actuator is shown in Figure 7. The cup design encloses the magnetic flux, hence avoiding magnetic crosstalk between actuators, which was an issue for Echidna.

The required pitch for Mohawk is 6.75mm at the focal surface. The focal surface is convex with 8m ROC, and the pitch at the modules is a few percent less than this, depending on the spine length. The WFMOS-A actuator and piezo design cannot be used on a pitch less than ~8.3mm, so modification as required for Mohawk. Mohawk requires a smaller ball bearing and piezoceramic tube, and a thinner spine tube. For this design, we use a 6.35mm (1/4") ball, 4.5mm OD piezo, and 1.5mm OD spine. The Mohawk actuator design is shown in Figure 7.

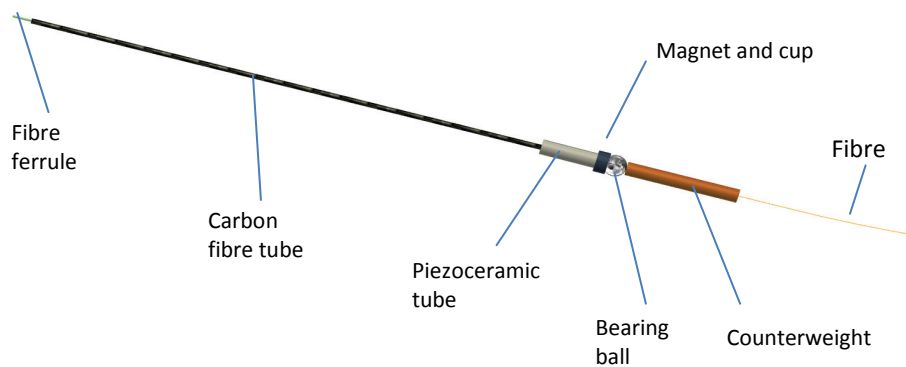

**Figure 7. Design for Mohawk actuator.**

The requirements for identical modules, identical spines, and a curved focal plane, together imply we must have curved modules, mounted in a fanned arrangement, like the staves of a barrel. The required sag is only 3.2 mm over the 450mm active length of the modules, and the fanning only amounts to deviations from parallel amounting to tens of microns per module. With this design, the tips of the spines, at their home positions, follow the required spherical surface and are telecentric.

There are 38 modules in total, fitted onto a frame, ensuring they are aligned and fanned correctly. Individual modules are acceptably stiff against gravitational flexure [10], but they will likely be clamped together laterally within the frame, to increase the overall stiffness further.

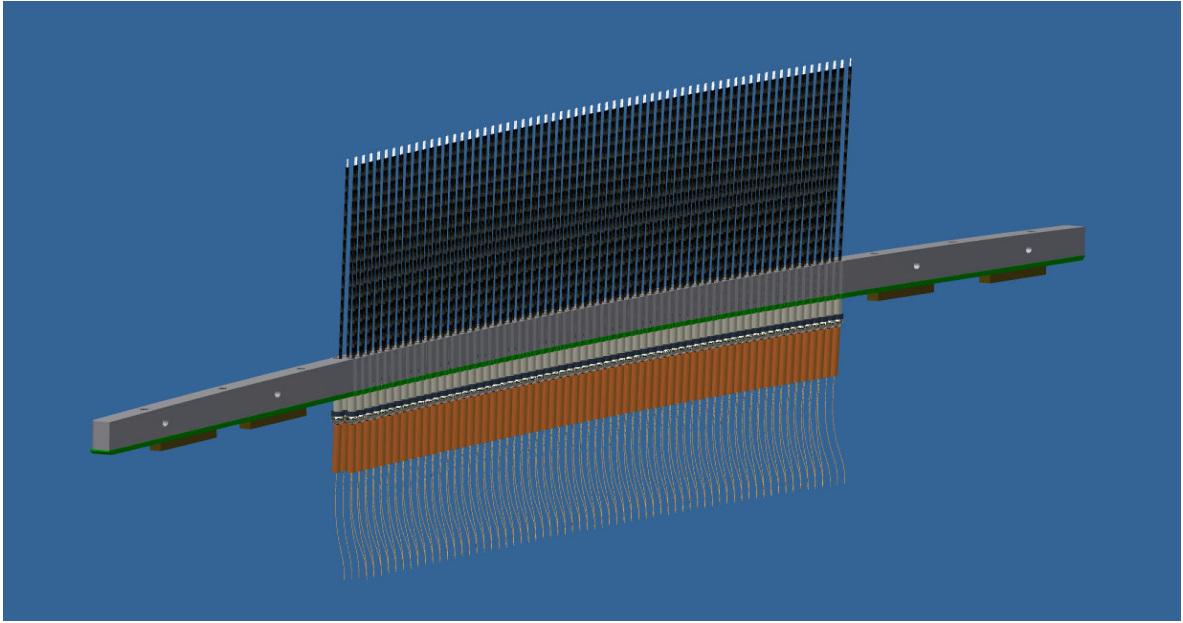

**Figure 8. Complete curved module fully populated with actuators, also showing the multilayer circuit board (green) and electrical sockets (brown).**

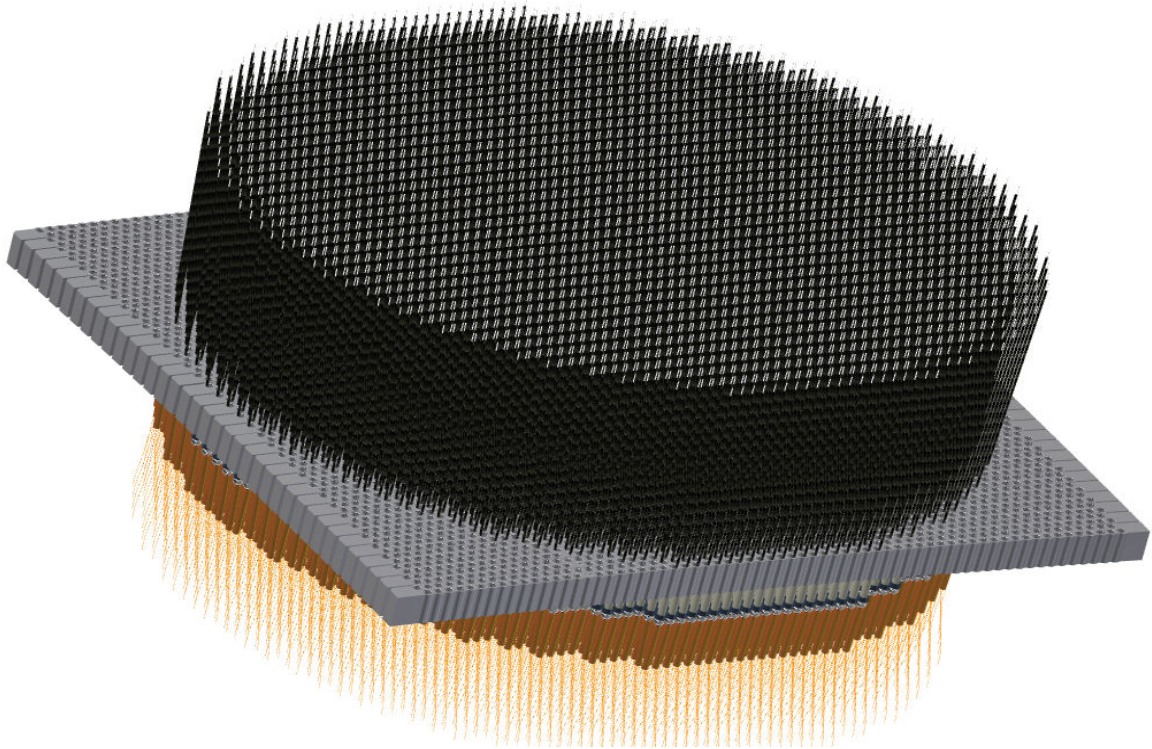

**Figure 9. Complete Mohawk actuator array, with 4000 spines covering the 450mm diameter curved focal surface**

The patrol radius is limited by the OD of the carbon-fiber spine tube, the ID and length of the piezoceramic tube, and the diameter and length of the hole through the module base. However, these quantities are all constrained by the needs for functionality and rigidity. To maximize the patrol radius, the modules have 'crinkle cut' sides (Figure 10). This also allows individual modules to be removed from the instrument without disturbing the neighbouring modules.

Power is brought to the actuators by two multilayer circuit boards (one from each end), glued to the module. The number of layers is greater than previous designs (~35 instead of 21 for Echidna), but the minimum thickness of individual layers has reduced over the years, and the overall thickness is much the same. This board would need to be curved by a few mm: this needs to be tested but is not expected to be an issue.

Out of interest for future instruments, we investigated the minimum pitch for this technology, and pitches of 6mm or less look plausible. The limiting factors are the track-widths and the gaps between them on the circuit board, the increasing number of PCB layers as the number of actuators/module increases, and the minimum reliable step size, which correlates inversely with ball bearing size. A smaller pitch also has some advantages, in that it reduces the focal and telecentricity errors, and reduces the required voltages to 100V or less.

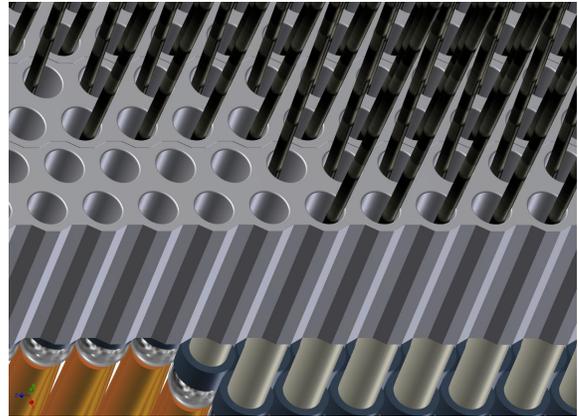

**Figure 10. Close up of 3 adjacent modules.**

## 3. PROTOTYPING

A 6-actuator prototype 'Minihawk' closed-loop positioner has been built at AAO (Figure 11, video clips 1 and 2). The design is not identical to that proposed for Mohawk, but the differences are minor. The spine-length is 160mm. This positioner is capable of positioning all 6 spines to within 7μm accuracy within 15s; and with improved calibration (i.e. accurately determining the step size and relative orientation for each fiber), this positioning time should decrease.

We have also experimented with increased spine length on individual spines, and up to 250mm does not provide any apparent additional problems. We have therefore taken 250mm as the nominal spine length for the Mohawk design.

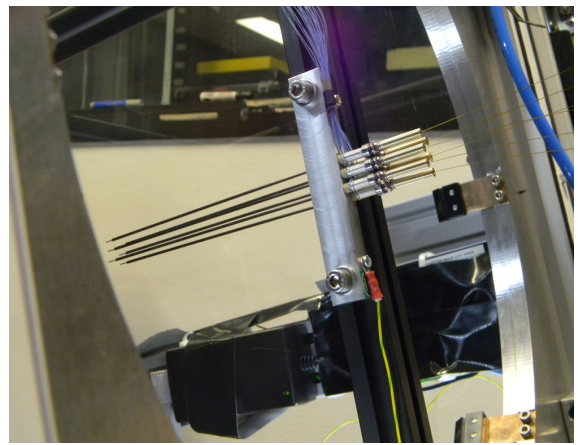

**Figure 11. 6-actuator 'Minihawk' Prototype, mounted in the AAO rotatable testing rig.**
**Video 1. Minihawk calibration sequence.**
http://dx.doi.org/doi.number.goes.here
**Video 2. Minihawk positioning sequence.**
http://dx.doi.org/doi.number.goes.here

## 4. ELECTRONICS

Each module will have two multilayer PCB boards, meeting half-way along, and glued to the module (Figure 12). Each board brings power to one half of the actuators on the module. The best option to get the power onto and off the curved and fanned modules has not yet been decided, but there are several viable options. The simplest would be to curve the master board into a cylinder with the appropriate ~8m ROC (again just a few mm of sag), and mount the sockets on the modules at the appropriate angle to make them parallel with the master board. Figure 13 shows the proposed arrangement. Various other options are possible - mount the required connectors on flat master boards at the appropriate heights and angles, use flexible circuit boards, or ribbon cables. Each module is then connected to a switching card, which controls the duration, orientation and polarity of the sawtooth signal to each actuator. These switching cards are mounted on 4 master boards, each of which controls one quadrant of the actuators. Figure 14 shows the proposed electronics layout.

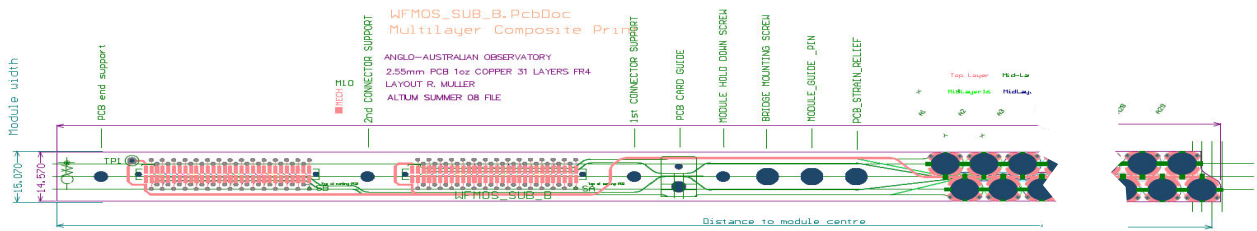

**Figure 12. Module PCB layout for WFMOS. Mohawk will be almost identical except for more actuators and bigger sockets.**

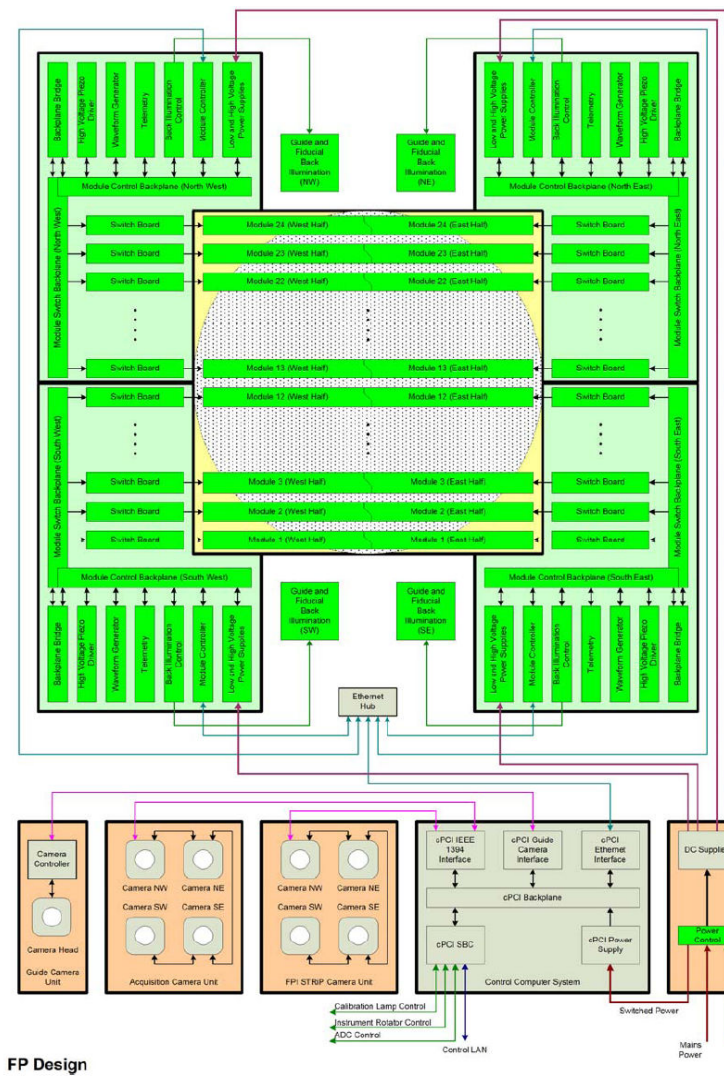

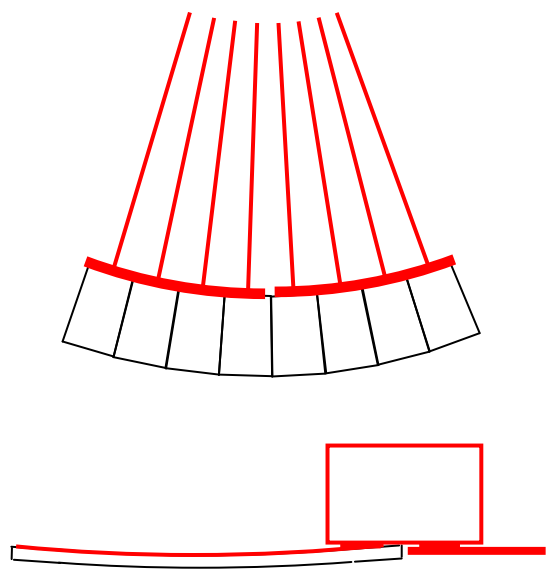

**Figure 13 (a) Schematic end view of assembled modules showing in red the fanned vertical switching boards, and the master boards, seen end on. (b) Schematic side view of a single module, showing vertical switching board, the section through the master board, and the sockets between module and vertical switching board, and between vertical switching board and master board. Individual actuators and spines are not shown.**

**Figure 14. Proposed electronics layout**

The heat load of the instrument has been scaled from earlier, more detailed studies. During actual repositioning, i.e. for some tens of seconds per hour, the heat dissipated would be 2-5kW. At all other times the power consumption would be a few hundred watts. It is expected that the instrument will be cooled, using the cooling water/glycol 15°C system already in place for DECam.

The back-illumination system also creates heat, ~ 0.1W/fiber, or 40-60W/spectrograph, but only when the LEDs are 'flashed', for <<1s each time, whenever metrology is performed (~5 times per repositioning move). The average heat load is then less than 0.1W/spectrograph and can be ignored.

## 5. FIBERS AND CONNECTORS

The system will likely have to be connectorised somewhere close to the positioner, to allow instrument changes[5]. Tests at AAO have shown that standard telco MTP connectors give very good performance, when combined with Nye OC431A optical coupling gel, with throughput losses amounting to just a few percent or less [6]. The gel and the requirement for cleanliness means reconnecting is a significant job, but still routinely doable by one person in half a day.

---

[5] The modular design of the positioner, means that designs without connectors can be considered. Each spectrograph would be fed by a fixed set of 5 - 10 modules, internally clamped together in normal use, and removed and replaced together. Each spectrograph would have its own fiber cable, and this would contain spare fibers for repairs. These cables would be strapped together in normal use. The slit units would be detachable from the spectrographs, and boxes would be supplied allowing both the slit units and the positioner modules to be safely stowed and removed along with the fiber cable. 2dF has such a box for the slit units, and the whole cable and box has been removed and replaced from the AAT top end to the coude room scores of times, without any mishap in 6 years.

Each connector has 24 fibers, and they can be bought in blocks of 4. 4-6 such blocks would connectorise each spectrograph, and ~40 blocks of connectors would be needed in total. Standard MTP connectors are designed for fibers with 125μm core+cladding diameter, so it makes sense to design for this if we can.

At least two good choices of actual fiber are available, CeramOptic Optran WF or Polymicro FPB. Both have excellent transparency and Focal Ratio Degradation (FRD) performance. They have Numerical Aperture NA=0.22, safely faster than the expected collimator acceptance speed of ~F/2.75 (see below).

A cladding thickness of 10λ is required to avoid evanescent losses, so our maximum core diameter is then ~105μm. The preforms (cylindrical glass blocks from which the fibers are drawn) are now available in a wide range of standard core/cladding ratios, with 1:1.2, 1:1.25, 1:1.3 and 1:1.4 all being available from one or other manufacturer. This means that almost any fiber core size in the range 89μm (1.56″) to 104μm (1.83″) can be accommodated, while maintaining the 125μm core+cladding diameter. While the latter is almost perfect in terms of maximizing S/N for sky-limited observations of faint galaxies, the former would allow slower transmissive cameras in the spectrographs, while still giving the required spectral resolution [7].

## 6. FOCAL RATIO DEGRADATION AND THROUGHPUT

The tilting spine design introduces unavoidable telecentricity and focus errors, but these reduce as the spine length increases. We will use the longest spines compatible with robust functionality, and this can only be determined by prototyping. For a spine length of 250mm (the current prototype), the maximum telecentricity error caused by the spine tilt is 1.55° (compared with the 9.8° half-width of the telescope beam). This causes 'geometric' FRD, with the beam being apodised on exit from the fibers. We can accurately model the effect as a convolution with a hollow conical beam of half-width 1.55°, and the effect of this is shown in Figure 15. The spectrograph collimator speed can be increased to collect this apodised light, but this decreases the demagnification between collimator and camera, and hence decreases spectral resolution and the number of fibers that can be accommodated per spectrograph. Therefore, we would expect a trade-off between collimator speed and throughput, and shown in Figure 15 is a nominal collimator speed of F/2.75. The maximum light lost, due to the telecentricity error introduced by spine tilt, then amounts to 6.3%.

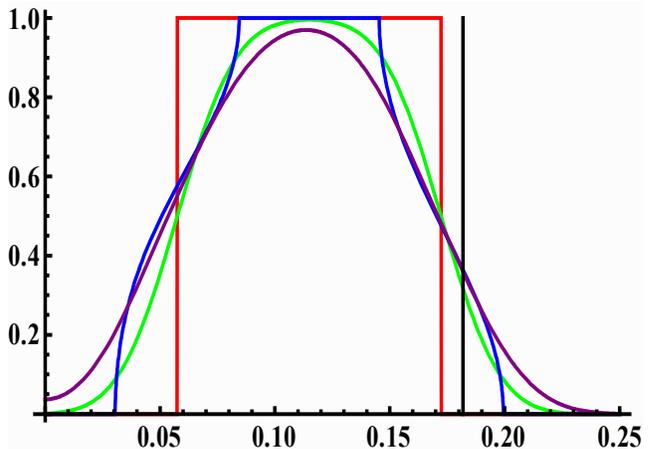

**Figure 15.** Intensity vs Numerical Aperture for an F/2.9 telescope beam with F/9 central obscuration (red), as affected by normal FRD (simulated as a convolution with a 2.7° FWHM Gaussian) (green), and by a 1.55° telecentricity error at fiber input (blue), and the combined effect (purple). The vertical black line corresponds to an F/2.75 collimator acceptance speed.

All fiber systems introduce 'normal' FRD in any case, just from deviations from perfect cylindrical geometry. However, the tilting-spine design is very gentle on the fibers, since there is no twisting, tension or even significant bending; this contribution to FRD should then be as good as can ever be achieved with fiber systems. FRD causes the far-field beam exiting the fibers to be apodised, and the effect is reasonably well approximated by a convolution of the telescope beam with a Gaussian beam with FWHM F/15-F/20[6]. The effect is also shown in Figure 14. The light lost, due to normal FRD, by this overfilling of the collimator amounts to 5.5%. The combined effect of both FRD losses is 10.0%.

For 250mm spine length, the difference in focus between zero and maximum tilt is 92μm. If the focus is set at one of these positions (see below), the maximum defocus at the other corresponds to additional PSF FWHM of 0.46″, compared with ~0.7″ for the telescope optics and ~0.75″ for the seeing. The increased aperture losses caused by this defocus, for a typical faint emission-line galaxy of FWHM 0.8″, is ~6.2%. The throughput losses caused by defocus and tilt can be balanced against each other, by setting the axial position of the positioner such that spines are in focus when close to their maximum patrol radius. Thus the defocus introduces no additional losses.

---

[6] Because the FRD adds an angular dispersion to the beam, the *fractional* degradation in etendue caused by FRD is minimized by using the fastest beam speed possible (i.e. compatible with the NA of the fibers). Slower beams actually incur larger FRD dispersion, even in fixed angular terms (when collimated light is injected axially into a fiber, the far-field beam on exit is an annulus, and the width of this annulus clearly declines with increasing input angle [e.g. 8]). This reinforces the argument above about optimal speeds within fibers, but means the infilling of the central obstruction is always greater than calculated here.

The only other throughput losses between aperture and collimator are from the fibers themselves, and connectors, if used. The fiber input surface can be AR-coated [9], and the fiber slit can also be in optical contact with an AR-coated field lens, to give <1% combined losses. The connectors introduce ~3% losses, to give a total loss attributable to the positioner system itself of ~15%. To this must be added the attenuation losses within the fibers themselves, and these are shown in [1]. The loss which is attributable to the tilting-spine technology is 6.3%.

## 7. TARGET AND FIBER YIELDS

We have set the physical parameters of the actuators by demanding that the patrol radius be equal to the pitch. This gives reasonable piezoceramic tube dimensions and adequate module stiffness (scaling from the FEA modeling done for WFMOS). However, during detailed design, we might be able to increase the patrol radius further.

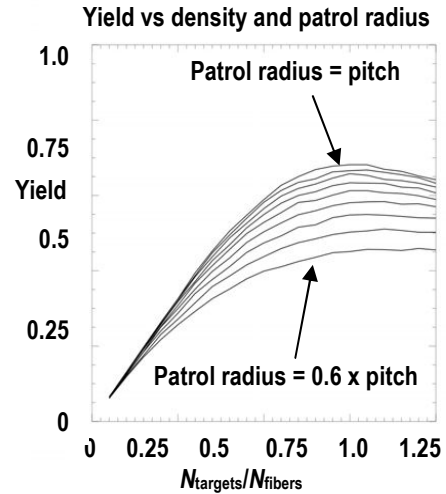

The 'fiber yield' is defined as the fraction of fibers assigned to science targets in a typical observation, and the 'target yield' as the fraction of potential science targets getting a fiber assigned to them. Both depend on the relative density of targets and fibers on the sky, on the patrol radius of the fibers, and on the clustering of the targets. For both faint galaxies and for field stars, we can neglect the clustering and use Poisson statistics. The resulting product of fiber and target yields, as a function of the relative densities and the patrol radius, is shown in Figure 16. For large patrol radii, this combined yield has a maximum when the target density is about equal to the fiber density. A larger patrol radius always implies a greater combined yield, but the gain diminishes as patrol radius increases. For patrol radius equal to the pitch, the combined completeness is ~0.70, implying that ~84% of fibers get used in each observation, and 84% of potential targets get observed[7].

**Figure 16. Combined yield vs relative density and patrol radius (in steps of 0.05 × pitch). Adapted from [10]**

This leaves 16% of the fibers for use either as dedicated sky fibers (used for measuring the sky spectrum, to subtract from object+sky spectra), or for use on other science targets. The optimal fraction of dedicated sky fibers is $\sim\sqrt{(M/N)}$, where $N$ is the number of fibers in each spectrograph, and $M$ is the number of significant degrees of freedom, both in the sky spectrum variations across the field, and in the Point Spread Function variations caused by spectrograph aberrations. For Principal Component Analysis (PCA) (or similar), $M$ is typically about 10. $N$ is expected to be 400-600. This means the number of sky fibers arising naturally is close to what is in any case optimal[8].

The minimum separation of targets is set by the ferrules containing the fibers at the tips of the spines. These are nominally 0.7mm OD. Smaller separations could be easily enough achieved by machining the ferrules to be tapered at their tips (as was done for Echidna), and so we have taken 0.5mm (9″) as the closest proximity of configured targets. This gives an additional 2% incompleteness for a single pass[6].

## 8. BACK-ILLUMINATION, POSITIONING TIMES, AND METROLOGY

Back-illumination is mandatory, to allow closed-loop positioning of the actuators. FMOS-Echidna includes a deployable fiber connector, allowing back-illumination without interfering with the spectrographs. For DESpec, even if connectorised somewhere, a system allowing back-illumination of the fiber slit would likely be preferable. This requires a light-source that can be deployed in front of the fiber slit. With a transmissive collimator, or a reflective collimator including a large fold mirror like VIRUS [e.g. 2], this is straight-forwardly done by having the back-illumination system on a slide, or by using a small deployable fold mirror. This small fold mirror can also, in principal, act as the system shutter. With a reflective system with no fold mirror, a rather larger deployable unit is required, and it is more difficult to avoid light-leakage to the detectors while back-illuminating. If a suitably light-tight system cannot be devised, then it would be impossible to reposition while reading the detector, which puts an increased premium on fast repositioning.

---

[7] For the DESpec survey, the sky will be covered twice over, and this can be used to mprove the overall completeness, by ensuring galaxies in clusters or close pairs are preferentially targeted on the first pass. The final incompleteness should then just be a few %.
[8] For the WiggleZ survey using AAOmega on the AAT, the emission-line galaxies observed (which are not as faint as those proposed for DESpec) are so much fainter than sky, that their continua are completely invisible within the sky noise. They are also at widely different redshifts within each field. This means that, paradoxically, the set of such object spectra makes an almost perfect set of sky spectra for input into PCA-type analysis. If this technique is adopted for DESpec, 16% of the fibers are then available for use on other targets.

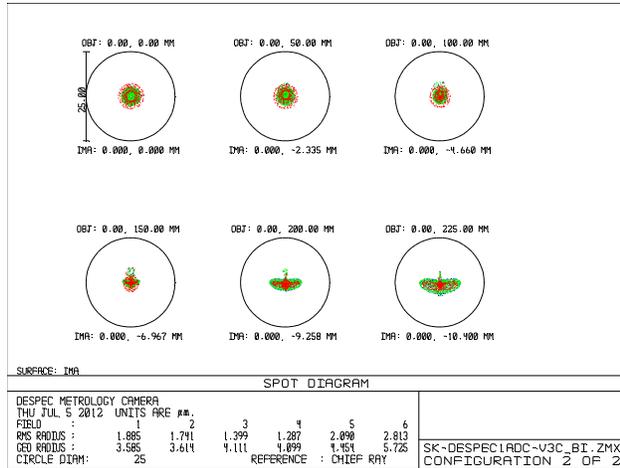 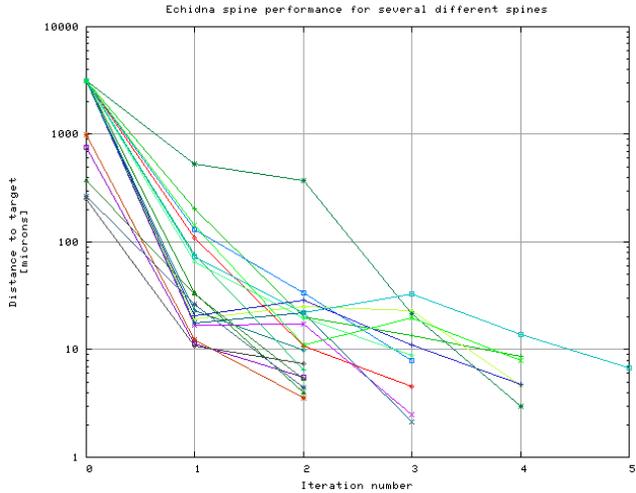

**Figure 17. Spot diagrams for the metrology system. Wavelengths are for red LED. Circle diameter is 25μm (corresponding to 0.5mm at focal surface). Airy circle is 8μm**

**Figure 18. Positioning error vs iteration number for Echidna. The Mohawk actuator design is significantly more repeatable, so eventual performance should be better than this.**

It is proposed that the metrology will use a long focal length camera, simultaneously imaging the entire focal plane, from a point within the Cassegrain chimney, on the telescope axis. The camera should contain a minimum of ~6K x 6K pixels to achieve the required metrology accuracy of ~5μm. Figure 17 shows the spot diagrams for such a system, using a 100mm aperture and just a pair of doublets, each with one weakly aspheric lens, imaging the focal surface onto a 21 × 28mm 120Mpix detector (as will be shortly available from Canon). The images are deliberately defocused, but they are still virtually diffraction-limited, and easily small enough to be resolved from each other when spines are at their nominal minimum separation of 0.5mm. The images are almost achromatic and coma-free, which simplifies the mapping from camera to physical coordinates.

This arrangement means the entire array of spines can be repositioned in parallel. Several iterations are typically required for accurate position (Figure 18), though overall repositioning time is determined by the most recalcitrant actuator. The estimated total repositioning time is estimated as 15s, with a very conservative upper limit of 30s.

To tie the spines to an absolute frame of reference, there will be 'fiducial' spines, fixed in position but otherwise identical to the science spines, mounted outside the circular focal plane, and which can be back-illuminated in the same way as the science and acquisition spines.

## 9.  ACQUISITION AND GUIDING

For guiding, guide spines would be provided, each containing a small coherent bundle of fibers, but otherwise identical to the science spines, so that flexure and thermal effects are the same in both. The individual apertures in the coherent bundles used for guiding will have a pitch ~1" (57μm) or less, to adequately sample the seeing. At least 6-8 guide stars, with V<15[m], are needed for each field (to give redundancy against incorrectly known proper motions), well distributed around the outer parts of the field. Because each guide spine has only a small patrol area, there needs to be a large number (20-100) of them, to be sure of finding enough guide stars. However, there are hundreds of spare slots in the outer modules, just outside the science field of view, but still receiving adequate, albeit vignetted, skylight[9].

---

[9] The progressive vignetting across the coherent bundle introduces a small but correctable error into the centroiding.

As part of the DECam upgrade, the Telescope Control System is being overhauled, and the pointing accuracy of the Blanco telescope is expected to be a few arcsec rms. The ID of the carbon fiber rods in a standard spine is 0.7mm, allowing a coherent bundle with diameter ~10" on the sky. If the telescope pointing is as good as hoped, no further acquisition system may be needed at all. However, for surety, and also for commissioning the instrument, it is very helpful to be able to image larger fields of view, to find bright stars to determine the precise plate-scale and orientation. For this purpose, a small number of acquisition spines are proposed, containing larger (~20″) coherent bundles at the expense of a smaller patrol radii. These acquisition spines would also allow fields to be quickly acquired, for any reasonable telescope pointing precision. Around 10 of these are proposed, including one at or near the center of the field.

Field rotation is required for both acquisition and guiding[10], but is already provided by the DECam hexapod.

## 10. TELESCOPE MOUNTING

DESpec must be interchangeable with DECam, on a timescale of a day or so. DESpec reuses the DECam WFC lens barrel, containing DECam lenses C1-C4. DESpec will contain two new lenses, C5' and C6, permanently mounted in front of the positioner. This allows a Mohawk to be a completely dust-sealed system. DESpec will be mounted on the flange at the end of the DECam WFC barrel, and will have the same arrangement at the rear to allow use of the DECam mounting jig positioned at the Blanco NW prime focus access position.

The available volume for DESpec is cylindrical, with diameter and depth both over a meter, comfortably large enough for the Mohawk positioner, and its electronics, and the two new corrector lenses. The overall layout is shown in Figure 19. The weight of Mohawk would be ~500kg, similar to the DECam camera it replaces.

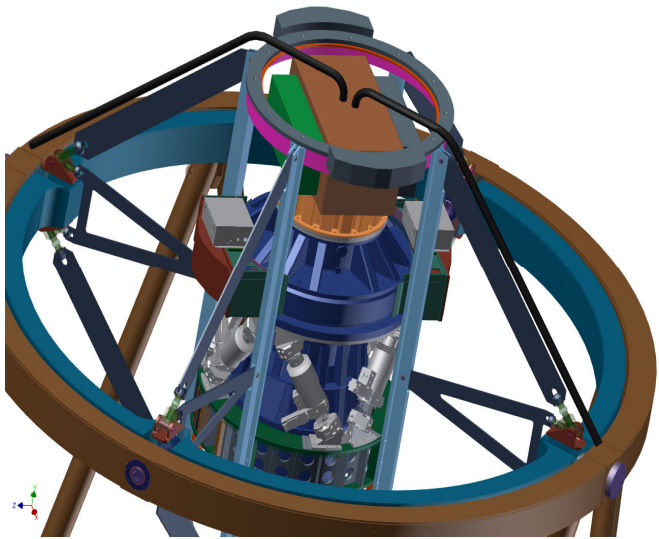

**Figure 19.** Mohawk mounted on the top end of the Blanco telescope, attached to the DECam Wide Field Corrector assembly (blue, with hexapod). The new lens assembly is in the orange mounting, Mohawk itself is in the brown box, with electronics in the green and grey boxes on its sides. The fiber cables are black.

---

[10] Field rotation is needed to correct for telescope polar axis misalignment, and to partially correct for changes in differential refraction during integrations.